# Attitude and Motivation towards Learning Physics

Prof. Ryan Manuel D. Guido
*Rizal Technological University*

## Abstract

*Physics is considered as one of the most prevailing and problematic subject by the students in the realm of science. Students perceived physics as a difficult subject during high school days and become more evasive when they reach college. Based on an investigation of 446 engineering and technology students, this study analyses and evaluates the relationship between attitude and motivation of engineering and technology students towards learning physics. It shows that there is no significant difference in the attitude and motivation of students towards learning physics. Furthermore, there is a negligible degree of relationship for attitude and motivation. The relationship between attitude and motivation is due to chance. This confide that most students found that it is unlikely for them to enjoy learning physics because of their professor. Faculty should encourage students in drilling physics problems, students are found to like answering difficult physics questions in every examination, expresses that they have a mathematical mind.*

## 1. Introduction

One of the most prevailing subjects in the field of engineering is physics. Its prominence in the academic and specialized range in engineering students which makes it a support of more than instrumental substance for learning all throughout the engineering course. It is the basic foundation of every engineering topics and the pedagogical significance to every engineering student.

Physics is considered as the most problematic area within the realm of science, and it traditionally attracts fewer students than other sciences like chemistry and biology. Most of the students perceived physics as a difficult subject during high school days and becomes more problematic when they are in college, and even more challenging in graduate education. In the Philippine setting, it is more perplexing to most of the Filipino students to study physics because of the undesirable reputation long before time. With this, only students who do well in high school physics and students that are exceptionally good in mathematics, remarkably talented and gifted in science can appreciate the role of physics in their daily life.

With such discourse in the physics education in the Philippines, students find it to have a negative attitude toward learning physics because of its computational exigency in every problem sets, moreover, if they don't like the subject more often they don't like the teacher. One of the most prevailing teaching-learning strategies and approaches for physics teachers is to make physics amusing and entertaining in the class. Physics can be applied by several teaching methods in order to transfer learning effectively and efficiently. Such effects in teaching physics may increase the ability of the students to understand and comprehend more than the usual and the ability of their learning process can be realized.

It was observed that the students who have negative attitude towards physics have lack of motivation for class engagement, and also the students who have positive attitudes towards physics have motivation for class engagement (Cracker, 2006 [7]).

## 2. Literature Review

### 2.1. Students' Attitude towards Learning

Attitude can distort the perception of information and affect the degree of their retention. Also, it affirmed that students' attitudes and interest could play substantial role among pupils studying science, and attitude implies a favorable or disfavorable evaluative reactions towards something, events, programmes, etc. exhibited in an individual's beliefs, feelings, emotions or intended behaviors. It also shows that students' positive attitudes to science correlate highly with their science achievement.

One of the utmost significant factors which affect students' academic success is their attitudes towards school, lessons and academic success. Attitude is a tendency for individuals who organize thoughts, emotions and behaviors towards a psychological object. Human beings are not born with attitudes; they learn them afterwards. Some attitudes are based on people's own experiences, knowledge and skills, and some are gained from other sources. However, the attitude does not stay the same, it changes in the course of time (Erdemir, & Bakirci, 2009 [9]).

Learner's motivation in learning is affected by their attitudes towards learning the subject. The relation between motivation and attitudes has been considered a prime concern in learning. According to Gardner and Lambert (1972) [13], motivation to learn is thought to be determined by his attitudes towards the other group in particular and by his orientation towards the learning task itself. Only when paired up with motivation proper attitudinal tendencies relate to the levels of student engagement in learning, and to attainment.

Gardner (1980) [12], elaborates attitude as the sum total of a man's instinctions and feelings, prejudice or bias, preconceived notions, fears, threats, and convictions about any specified topic.

On the study conducted by Shuib (2009) [28], students' attitudes revealed that most of students had positive attitudes towards the social value and educational status of learning, in addition, the findings showed the students' positive orientation toward the language learning. Ajzan (1988) [3], ponders attitude as a disposition to respond favorably or unfavorably to an object, person, institution or event.

The measurement of students' attitudes towards physics should take into account their attitudes towards the learning environment (Crawley & Black, 1992 [8]). The effect of student's attitude toward science is incredibly important, because in problem solving requires patience, persistence, perseverance and willingness to accept risk (Charles, Lester & O'Daffer, 1987 [6]).

Pintrich and Maehr (2004) [25] classifies students in three groups such as the ones who avoid failure, the ones who would like to satisfy their curiosity and the ones who want to get high marks. The study shows that students in classes, their motivation degrees and strategies are different. When students have positive attitudes, they show positive behaviors and they fulfil their academic necessities.

Eryilmaz, Yildiz & Akin (2011) [10] examined the relationship between attitudes of high school students towards physics laboratory and being motivated for class engagement or not. They concluded that students who have high-level motivation for class engagement have also positive attitudes towards physics laboratory. In contrast with this conclusion, students who have low-level motivation for class engagement have negative attitudes towards physics laboratory.

Achievement, motivation and student interest are influenced by positive and negative attitudes (Miller, Abraham, Cohen, Graser, Harnack, & Land, 1961 [22]). Additionally, it is found out that students with positive attitudes towards physics had positive attitudes towards their science teachers, science curriculum and science-classroom climate. Students' attitude towards science is more likely to influence the success in science courses than success in influencing attitude (Morse and Morse, 1995 [23]).

If students have negative attitudes towards science, they also do not like physics courses and physics teachers. Based on this premise, numerous studies have been conducted to determine the factors that affect the students' attitudes in science. There are basic factors including: teaching-learning approaches, the use of the presentation graphics, the type of science courses taken, methods of studying, intelligence, gender, motivation, attitudes, science teachers and their attitudes, self-adequacy, previous learning, cognitive styles of students, career interest, socioeconomic levels, influence of parents, social implications of science and achievement (Craker, 2006 [7]).

Many researchers believed that if students were allowed to demonstrate higher cognitive abilities through problem solving, either through a teacher-centered approach or a student-centered approach, their attitudes towards physics might be positively affected (Erdemir, 2009 [9]). Furthermore, it was concluded in his study that the poor student attitudes towards physics in the control group was due to the lack of information, lack of problem solving skills, lack of self-confidence, using a formula incorrectly and lack of acting like experts while they solve physics problems.

## 2.2. Students' Motivation towards Learning

Motivation, according to Gardner (2006) [14], motivation is a very complex phenomenon with many facets. This is because the term motivation has been viewed differently by different schools of thought. Brown (2000) [5], identified motivation as quite simply the anticipation of reward, he also asserts that motivation of learners often refer to a distinction between two types of motivation namely, instrumental versus integrative motivation.

With lack of attitudes and motivation in the physics course, most of the engineering graduates have poor background in physics, yet, taking pre-requisite subjects made them difficult to understand more in succeeding topics.

Students' motivation can be external or intrinsic. External motivation generally consists of recognition and praise for good work. In a college student, this might be in the form of sustainability of the scholarships, or good impression in the class and at home. Students' grades is one of the most prominent factor as their extrinsic goal orientation. While intrinsic motivation generally consists of an internal desire to learn about a specific topic. According to Vansteenkiste, Simons,

Lens, Soenens, Matos, & Lacante (2004) [30] students demonstrated with intrinsic motivation processed reading material more deeply, achieved higher grades, and showed more persistence than students with extrinsic motivation.

Learners' motivation has been widely accepted as a key factor which influences the rate and success of learning. There are many factors that might cause the students' low proficiency, one might attribute to students' motivation towards the subject. According to McDonough (1983) [21], motivation of the students is one of the most important factors influencing their success or failure in learning.

A better understanding of students' motivation and attitudes may assist curriculum and instruction designers to devise language teaching programs that generate the attitudes and motivation most conducive to the production of more successful learners.

According to Maehr and Midgley (1991) [19], motivation for class engagement is one of variables. Motivation has a significant role in teaching and learning. But today, according to motivational perspective, students are considered as individuals who are able to reach a decision by assessing possibilities and consequences that can transfer their aims into life and form meaning. Motivation to class engagement means that students want to engage the class activities if they have motivation.

Based on the study of Eryilmaz, Yildiz & Akin (2011) [10], on investigating or relationships between attitudes towards physics laboratories, motivation and amotivation for the class engagement. Shows that the most significant problem that teachers confront in physics lessons is that abstract or concrete subjects cannot be comprehended by students correctly or efficiently. A lot of techniques and methods are used to handle this situation. Application of any technique or method but students' attitudes towards the lesson and their knowledge that they gain previously are mainly effective in learning.

If the students do not have motivation to participate in the lesson, many cases get then bored, they cannot focus their attention on the subject, and they cannot establish any connection with the studies done in the school and real life. As a result of the students' being bored and reduction of their lesson attention, it can be seen the drop out of the students (Pintrich & Maehr, 2004 [25]).

### 2.3. Students' Learning

Educational research shows that promoting metacognition in the science classroom prompts students to refine their ideas about scientific concepts and improves their problem-solving success (Rickey and Stacy, 2000 [26]).

Most learning strategy theories are based on the constructivist perspective of learning which contends that meaning and knowledge are constructed by the learner through a process of relating new information to prior knowledge and experience (Olgren, 1998 [24]). It also emphasizes that the quality of learning outcomes depends on how well the learner organizes and integrates the information.

Zuscho, Pintrich and Coppola (2003) [32] investigated how students' level of motivation and use of specific cognitive and self-regulatory strategies changed over time, and how these motivational and cognitive components in turn predicted students' course performance. The result showed that there is an overall decline in students' motivational level over time. There was also a decline in students' use of rehearsal and elaboration strategies over time; students' use of organizational and self-regulatory strategies increased over time.

Deep approach intends to understand material for oneself, vigorous and critical interaction with knowledge content, relating ideas to one's pervious knowledge and experience, discovering and using organizing principles to integrate ideas, relating evidence to conclusions, and examining the logic of arguments (Knowles and Kerkman, 2007 [18]). For deep learning to occur, students should use a combination of organization and elaboration strategies to analyze and synthesize information in ways that build a mental model linked to prior knowledge in memory.

Studies reviewed suggest that there is a relationship between attitude and methods of instruction and between attitude and achievement. Therefore, it is possible to predict the level of achievement from attitude scores. Although many researchers argue that teaching methods have a great impact on students' attitude to learn a subject, students' attitudes towards physics have not yet been examined. In this study, the effects of the problem-solving strategy on students' attitudes were investigated (Eridemir, 2009 [9]).

### 2.4. Students' Attitude and Motivation in Science Education

It was observed that the students who have negative attitude towards physics have lack of motivation for class engagement, and also the students who have positive attitudes towards physics have motivation for class engagement. Craker (2006) [7] demonstrated that attitudes towards science change with exposure to science, but the direction of change may be related to the quality of that exposure, the learning environment, and teaching method.Similar results were obtained in the study conducted by Mattern and Schau (2002) [20] after exposing students to a self-learning device.

On the study conducted by Seery (2009) [27], he recommended that relevant stake holders to continue promoting positive attitude towards chemistry and motivate students to work hard in order to achieve better in the subject as it cuts across all science related careers. Aiken (2000) [2] relates that attitude affects people in everything they do and in fact reflects what they are, hence a determining factor of students' behaviour.

Wilkinson and Maxwell (1991) [31] affirmed that problem solving using ill-structured problems motivates students and encourages understanding the epistemology of the discipline.

On the study conducted by Haussler and Hoffman (2000) [17], regarding the development, comparison with students' interests, and impact on students' achievement and self-concept revealed that as a result of the comparison between the features of the curricular frame, the interest structure of students, and the current physics curriculum, has a remarkable congruency between students' interest in physics and the kind of physics education identified in the Delphi study as being relevant. However, there is considerable discrepancy between students' interest and the kind of physics instruction practiced in the physics classroom. It also appears in the Regression analysis that students' interest in physics as a school subject is hardly related to their interest in physics, but mainly to the students' self-esteem of being good achievers.

### 2.5. Physics Education

In the study of conventional teaching and traditional teaching methods, results shows that in order to increase the level of attitude and success in physics education, new teaching methods and technology need to be implemented into physics education (Adesoji, 2008 [1]).

There are several teaching methods can be used in teaching physics. Problem solving is one approach that is most in practice due to its mathematical facets of the concepts. In this approach, it involves knowing what to do in the situation of not knowing what to do. Problem solving is not only finding the correct answer, but also is an action which covers a wide range of mental abilities. Students should realize what and why they are doing, and know the strengths of these strategies, in order to understand the strategies completely and be able to select appropriate ones (Erol, Selcuk & Calishan, 2006 [11]).

A positive attitude influences expected achievement and is heavily influenced by attitudes towards science. As would be expected, positive heavily influenced by attitudes towards science also lead to better results on achievement influencing attitude (Craker, 2006 [7]).

Knowledge and skills related to solving physics problems are essential to ensure a positive attitude toward physics. This result agrees with Tooke and Lindstorm's (1998) [29] opinion that students who have a positive attitude towards and beliefs about physics will succeed at a higher level.

On the study conducted by Guido and De la Cruz (2012) [15], shows that the astronomy students are very optimistic about their academic performance. Exploratory research has revealed the reason associated with students' attitudes towards physics courses and methods of teaching (Craker, 2006 [7]). It is also expressed that pleasures in physics course if the students know how to plan and implement the strategies of solution to the question through teaching methods.

Physics education research has demonstrated that knowledge and skill of students taught problem solving techniques or methods can play a powerful role in how they use the knowledge they have learned in their physics courses. Student attitudes towards physics also play a powerful role in how they think about using problem-solving method in their physics or any science class (Hammer, 1996 [16]).

## 3. Methodology

### 3.1. Research Method

The research indicates the inherent characteristics with the gathering of data. It purports and presents facts concerning the Attitude and Motivation Scale of Physics Students. It also involves in the analysis on the assessment of the result of data gathered from questionnaires distributed to the respondents. It gives significance to the quality and standing facts.

A survey technique correlational research design measures variables for a large number of cases to see relationship. It is a procedure in which subjects' scores on two variables are simply measured, without manipulation of any variables, to determine whether there is a relationship. The study also involves selecting subjects when it is very difficult to locate which shows characteristics of the study and uses the fastest way of getting their opinion, reaction, responses to the Instrument in Attitude Inventory Test and Achievement Motivation Scales of the physics students.

### 3.2. Participants of the Study

The samples are 290 engineering students, and 156 technology students taking up engineering physics during the second semester of 2011-2012 in Rizal Technological University. This study is delimited on the Department which has both Engineering and Technology offerings; these are Civil Engineering and Technology, Computer Engineering and Technology,

Electrical Engineering and Technology, Electronics and Communications Engineering and Technology, Instrumentation and Control Engineering and Technology and Mechanical Engineering and Technology.

### 3.3. Instrument

The instrument used standardized Attitude Inventory Test and Achievement Motivation Scale to measure the achievement of the students. It is designed and patterned to obtain adequate information regarding the attitudes and motivations of the students. The Attitude Inventory Test and Motivation Scale survey questionnaire has five degrees of intensity with weights of 5 being the highest and 1 being the smallest rating. The scale of the statistical value adopted to assess their attitude and motivation is as follows:

| Weighted Mean | Arbitrary Value | Verbal Interpretation |
|---|---|---|
| 4.51 – 5.00 | 5 | Very True to Me |
| 3.51 – 4.50 | 4 | True to Me |
| 2.51 – 3.50 | 3 | Somewhat True to Me |
| 1.51 – 2.50 | 2 | Untrue to Me |
| 1.00 – 1.50 | 1 | Very untrue to Me |

### 3.4. Research Procedure

The evaluation of the students' learning toward physics subject has two parts: one through Attitude Inventory Test and the other which entails the Achievement Motivation Scale.

The assessment of 446 physics students' responses toward the survey questionnaire in Attitude Inventory Scale and Achievement Motivation Scale yielded a scientific investigation. The respondents were given questionnaires in their physics class to identify their assertiveness and enthusiasm towards their physics subject. This was done at Rizal Technological University in the College of Engineering and Industrial Technology taking the physics subject second semester 2011- 2012.

### 3.5. Data Analysis

Student responses to the instrument were coded based on the 5-point Likert Scale determination per criteria so that higher scores embodied more positive responses. Using SPSS program, a one-way analysis of variance (ANOVA) was conducted to determine the significant difference in the attitudes and motivation. Pearson's Product Moment Correlation to determine the significant relationship between the attitudes and motivation of the respondents.

## 4. Results and Discussion

### 4.1. Significant Difference in the Attitudes of the Respondents in Physics Subject

**Table 1**
**Analysis of Variance on the Attitudes of the Respondents in Physics Subject**

| Source of Variation | SS | df | MS | F | P-Value | F Crit |
|---|---|---|---|---|---|---|
| Between Groups | 0.006 | 1 | 0.006 | 0.038 | 0.845 | 4.098 |
| Within Groups | 5.954 | 38 | 0.156 | | | |
| Total | 5.960 | 39 | | | | |

*Level of significance at 0.05*

Table 1 reveals the ANOVA on the attitudes of the respondents in physics subject. This shows that the tabular value of 4.098172 is greater than the computed value of 0.038304 at correlation significance at 0.05 levels. The approximate significance is 0.845877. This indicates that it failed to reject the null hypothesis and therefore concludes that there is no significant difference in the attitudes of the respondents in the physics subject.

Studies reviewed suggest that there is a relationship between attitude and methods of instruction and between attitude and achievement. Therefore, it is possible to predict the level of achievement from attitude scores (Eridemir, 2009 [9]).

### 4.2. Significant Difference in the Motivation of the Respondents in Physics Subject

**Table 2**
**Analysis of Variance on the Motivation of the Respondents in Physics Subject**

| Source of Variation | SS | df | MS | F | P-Value | F Crit |
|---|---|---|---|---|---|---|
| Between Groups | 0.422 | 1 | 0.422 | 1.163 | 0.287 | 4.098 |
| Within Groups | 13.78 | 38 | 0.362 | | | |
| Total | 14.20 | 39 | 14.20 | | | |

*Level of significance at 0.05*

Table 2 reveals the ANOVA on the motivation of the respondents in physics subject. This shows that the tabular value of 4.098172 is greater than the computed value of 1.16395 at correlation significance at 0.05 levels. The approximate significance is 0.287449. This indicates that it failed to reject the null hypothesis and therefore concludes that there is no significant difference in the motivation of the respondents in the physics subject.

It was observed that the students who have negative attitude towards physics have lack of motivation for class engagement, and also the students who have positive attitudes towards physics have motivation for class engagement. Research has shown that attitudes towards science change with exposure to self-learning strategies in science (Craker, 2006 [7]; Mattern and Schau, 2002 [20]).

### 4.3. Significant Difference between Attitude and Motivation of the Respondents towards Learning Physics

Table 3
**Analysis of Variance on the Attitude and Motivation of the Respondents in Physics Subject**

| Source of Variation | SS | df | MS | F | P-Value | F Crit |
|---|---|---|---|---|---|---|
| Between Groups | 0.845 | 3 | 0.281 | 1.085 | 0.360 | 2.724 |
| Within Groups | 19.741 | 76 | 0.259 | | | |
| Total | 20.587 | 79 | 20.58 | | | |

*Level of significance at 0.05*

Table 3 reveals the ANOVA on the attitudes of the respondents in physics subject. This shows that the tabular value of 2.724944 is greater than the computed value of 1.08549 at correlation significance at 0.05 levels. The approximate significance is 0.360449. This indicates that it failed to reject the null hypothesis and therefore concludes that there is no significant difference in the attitudes and motivation of the respondents in the physics subject.

Similar results were obtained in the study conducted by Mattern and Schau (2002) [20] after exposing students to a self-learning device.

### 4.4. Significant Relationship between Attitude and Motivation of the Respondents towards Learning Physics

Table 4
**Pearson Product Moment Correlation on the Attitude and Motivation of the Respondents in Physics Subject**

| Correlation | Attitude | Motivation |
|---|---|---|
| Attitude | 1 | |
| Motivation | 0.923852 | 1 |

Table 4 shows the Correlation of attitude and motivation of the respondents towards learning physics. There is a high to very high positive relationship between the attitude and motivation of the students. The relationship is positive; as the attitude increases, the motivation also increases. The value of r = 0.923852 indicates a high to very high positive relationship between the attitudes and motivation of the 446 physics students at correlation significance at 0.05 level. The coefficient of determination is $r^2$ = 0.85350252, this presents that attitude improves prediction of the rated factors by 85.35%. The relationship between attitude and motivation is statistically significant.

This result agrees with Tooke and Lindstorm's (1998) [29] and Cracker (2006) [7] that positive attitude influences expected achievement and the opinion that students who have a positive attitude towards and beliefs about physics will succeed at a higher level.

## 5. Conclusion

The findings of the study suggest that most of the students find that they feel good when they are successful in physics. They feel that they are fully succeeded in the subject when this endeavor became fruitful. Factors that reflect on this is that their professors explain a lot per detail in their class, they also found it enjoyable studying because they found it useful for problems of everyday life. This also confirms that there is no significant difference in the attitudes of both engineering and technology students, the respondents' motivation for both samples; and attitude and motivation of the respondents in the physics subject. Moreover, the relationship between attitude and motivation is due to chance.

## 6. Recommendation

Most of the students found that it is unlikely for them to enjoy learning physics because of their professor. Faculty should deliver more and efficient teaching because students are seeking more than a satisfactory grade to pass the course, and they do not expect to have a lower grade in physics. Faculty should encourage students in drilling physics problems, students are found to like answering difficult physics questions in the exam, expresses that they have mathematical mind. The department should continue in motivating the students to learn more and study more as they expresses that they can get good results in physics, that they enjoy going beyond the assigned work in the subject and they try to solve more than what is expected of them, and that they found that studying physics is enjoyable.